\documentclass[a4paper,12pt]{article}
\pdfoutput=1
\usepackage{graphicx, rotating}
\usepackage{hyperref}
\usepackage{slashed}
\usepackage{ifpdf}

\ifx\pdfoutput\undefined
\usepackage[dvips,bookmarks=false]{hyperref}	
\else
\usepackage{hyperref}	
\fi
\hypersetup{colorlinks,bookmarksopen,bookmarksnumbered,citecolor=verdes,
linkcolor=blus,pdfstartview=FitH,urlcolor=rossos}
\def\hhref#1{\href{http://arxiv.org/abs/#1}{#1}} 

\usepackage{multicol}
\usepackage{color}
\definecolor{rosso}{cmyk}{0,1,1,0.4}
\definecolor{rossos}{cmyk}{0,1,1,0.55}
\definecolor{rossoc}{cmyk}{0,1,1,0.2}
\definecolor{blu}{cmyk}{1,1,0,0.3}
\definecolor{blus}{cmyk}{1,1,0,0.6}
\definecolor{bluc}{cmyk}{1,1,0,0.1}
\definecolor{verde}{cmyk}{0.92,0,0.59,0.25}
\definecolor{verdec}{cmyk}{0.92,0,0.59,0.15}
\definecolor{verdes}{cmyk}{0.92,0,0.59,0.4}

\font\tenrsfs=rsfs10 at 12pt
\font\sevenrsfs=rsfs7
\font\fiversfs=rsfs5
\newfam\rsfsfam
\textfont\rsfsfam=\tenrsfs
\scriptfont\rsfsfam=\sevenrsfs
\scriptscriptfont\rsfsfam=\fiversfs
\def\mathscr#1{{\fam\rsfsfam\relax#1}}

\newcommand{\riga}[1]{\noalign{\hbox{\parbox{\textwidth}{#1}}}\nonumber}

\newcommand{\eq}[1]{~{\rm (\ref{eq:#1})}}

\newcommand{\GeV}{\,{\rm GeV}}

\def\circa#1{\,\raise.3ex\hbox{$#1$\kern-.75em\lower1ex\hbox{$\sim$}}\,}

\newcommand{\eqref}[1]{(\ref{#1})}

\newcommand{\beq}{\begin{equation}}
\newcommand{\eeq}{\end{equation}}

\def\identity{1\!{\rm l}}
\def\eq#1{eq.~(\ref{#1})}
\def\circa#1{\,\raise.3ex\hbox{$#1$\kern-.75em\lower1ex\hbox{$\sim$}}\,}
\makeatletter

\def\art{\@ifnextchar[{\eart}{\oart}}
\def\eart[#1]#2#3#4#5#6{{\rm #2}, {#3 #4} {\rm (#6) #5} [arXiv:\-{\hhref{#1}}]}
\def\hepart[#1]#2{{\rm #2, arXiv:\-\hhref{#1}}}
\newcommand{\oart}[5]{{\rm #1}, {#2 #3} {\rm (#5) #4}}

\newcounter{alphaequation}[equation]
\def\thealphaequation{\theequation\hbox to
0.6em{\hfil\alph{alphaequation}\hfil}}
\def\eqnsystem#1{
\def\@eqnnum{{\rm (\thealphaequation)}}
\def\@@eqncr{\let\@tempa\relax \ifcase\@eqcnt \def\@tempa{& & &} \or
\def\@tempa{& &}\or \def\@tempa{&}\fi\@tempa
\if@eqnsw\@eqnnum\refstepcounter{alphaequation}\fi
\global\@eqnswtrue\global\@eqcnt=0\cr}
\refstepcounter{equation} \let\@currentlabel\theequation \def\@tempb{#1}
\ifx\@tempb\empty\else\label{#1}\fi
\refstepcounter{alphaequation}
\let\@currentlabel\thealphaequation
\global\@eqnswtrue\global\@eqcnt=0 \tabskip\@centering\let\\=\@eqncr
$$\halign to \displaywidth\bgroup \@eqnsel\hskip\@centering
$\displaystyle\tabskip\z@{##}$&\global\@eqcnt\@ne
\hskip2\arraycolsep\hfil${##}$\hfil& \global\@eqcnt\tw@\hskip2\arraycolsep
$\displaystyle\tabskip\z@{##}$\hfil
\tabskip\@centering&\llap{##}\tabskip\z@\cr}
\def\endeqnsystem{\@@eqncr\egroup$$\global\@ignoretrue} \makeatother

\newcommand{\SU}{\rm SU}

\begin{document}

\begin{center}
IFUP-TH/2010-07\hfill
CERN-PH-TH/2010-58

\bigskip\bigskip\bigskip

{\huge\bf\color{black} 
The electron and neutron EDM from\\[3mm]
supersymmetric see-saw thresholds}\\

\medskip
\bigskip\color{black}\vspace{0.6cm}
{\bf
\large Gian F. Giudice$^a$, {\bf Paride Paradisi}$^b$ {\rm and} \bf Alessandro Strumia$^{ac}$}
\\[7mm]
{\it $^a$ CERN, PH-TH, CH-1211, Geneva 23, Switzerland}\\
{\it $^b$ Physik-Department, Technische Universit\"at M\"unchen, D-85748 Garching, Germany} \\
{\it $^c$ Dipartimento di Fisica dell'Universit{\`a} di Pisa and INFN, Italia}

\bigskip\bigskip\bigskip\bigskip

{
\centerline{\large\bf Abstract}
\begin{quote}

\medskip

We consider the corrections that arise at one loop when integrating out heavy fields in supersymmetric models. 
We show that, in type-I see-saw models, complex $A_N$- and $B_N$-terms
of the heavy right-handed neutrino give radiative contributions to the neutron EDM, as well as new dominant contributions to the electron EDM. Type-II and type-III see-saw also predict a pure
gauge correction that makes complex the masses of the weak gauginos. All the see-saw models can predict observable EDM for the electron and for the neutron in a peculiar ratio.

\end{quote}}

\end{center}

\section{Introduction}
The observed neutrino masses give solid indications for the existence of some new physics at a superheavy scale 
$M\circa{<}10^{14}\GeV$. Identifying the nature of the dynamics at the scale $M$ is a very difficult task because 
experiments measuring neutrino masses and mixings probe only a dimension-five operator with coefficient suppressed by 
$1/M$. Given the largeness of the scale $M$, any other effect is essentially invisible to every experimental search, 
and the only possible handle is some indirect information from the requirement of a successful leptogenesis~\cite{review}.

Supersymmetric theories offer opportunities to probe experimentally some of the features of the physics at the scale 
$M$ --- which will be called here ``see-saw sector" --- because such dynamics could leave indelible traces on the soft 
terms of the light sparticles. The vestiges of the see-saw sector left on the soft terms correspond to interactions not 
suppressed by inverse powers of $M$, and therefore can lead to sizable and measurable effects. In particular, since the 
dynamics at the scale $M$ is responsible for neutrino masses and --- possibly --- for leptogenesis, violations of 
individual lepton number and of CP are expected in the soft terms.

The information about individual lepton number and CP violation is communicated from the see-saw sector to the supersymmetric 
Standard Model (SM) by renormalization-group (RG) effects above the scale $M$~\cite{Hall:1985dx,Borzumati:1986qx} and by 
finite threshold effects at the scale $M$~\cite{Farzan:2003gn}. In this paper we revisit these effects using the method of 
analytic continuation into superspace~\cite{Giudice:1997ni,ArkaniHamed:1998kj}. Besides simplifying the calculation, this 
method allows for a more transparent interpretation of the contributions to the soft terms obtained from integrating out the 
dynamics at the scale $M$.

But the most important result of our study is to reveal new effects that were not noticed
in previous analyses. 
Finite threshold corrections at the scale $M$ affect not
only the soft terms for the charged sleptons, but also the $A$ term for up-type quarks
as well as the $B_{\mu}$ term of the Higgs mixing. 
This latter term is also affected by the usual RGE running above $M$:
going to the standard phase convention where  $B_\mu\mu$ is real,
the $\mu$ term has a phase opposite to $B_\mu$.
This means that, if the $B$ and/or $A$-term of the right-handed neutrinos is complex, the see-saw sector gives comparable contributions to
{\it both} the electron and the neutron electric dipole moments (EDM).

In section~\ref{SS} we describe the calculation of the contributions to the soft terms
from integrating out the see-saw sector at the scale $M$ with the formalism of analytic continuation into superspace. We will consider the usual see-saw with singlet right-handed neutrinos (type-I) as well as type-II and type-III  see-saw mechanisms involving weak
triplets~\cite{review}. The integration of weak triplets at the scale $M$ gives pure
gauge threshold effects (namely independent of the unknown neutrino Yukawa couplings)
contributing to the gaugino masses and possibly inducing new CP violating effects.
The phenomenological implications of our results for the electron and neutron EDM are
discussed in section~\ref{EDM}. In section~\ref{end} we conclude summarizing our results.

\section{Threshold effects from superspace analyticity}\label{SS}

\subsection{Type-I supersymmetric see-saw}

The method of analytic continuation into superspace~\cite{Giudice:1997ni,ArkaniHamed:1998kj}
 allows us to extract information on the supersymmetry-breaking soft terms 
from calculations in the exactly supersymmetric theory, where we can fully exploit the power of the non-renormalization 
theorem. 

We start by considering the simplest see-saw sector, which contains one singlet right-handed neutrino $N$ for each generation. In the limit of exact supersymmetry, the theory is described by the Lagrangian 
\beq
\mathscr{L} = \int d^4\theta \left(N^\dagger Z_N N +L^\dagger Z_L L +H_u^\dagger Z_{H_u} H_u \right) +\left[ \int d^2\theta \left( N^T \frac{M}{2} N +N^T \lambda_N L H_u \right) +{\rm h.c.} \right] .
\label{susyL}
\eeq
Here the wave functions $Z_N$ and $Z_L$, the coupling constant $\lambda_N$, and the right-handed neutrino mass $M$ are $3\times 3$ matrices in generation space. In this scheme, the neutrino mass matrix is given by the matrix product
$m_\nu = - {\langle H_{\rm u}\rangle}^2 \lambda_N^T {M}^{-1} {\lambda}_N$. 
For the moment we neglect gauge interactions, which will be introduced only when necessary.

Supersymmetry breaking is described by the terms in the Lagrangian
\beq
\mathscr{L}_{\rm soft} =-N^\dagger m_N^2  N -L^\dagger m_L^2  L- {H_{\rm u}}^\dagger m_{H_{\rm u}}^2{H_{\rm u}}+\left( N^T \frac{B_NM}{2}N+N^T A_N \lambda_N  LH_{\rm u}  +{\rm h.c.}\right) ,
\eeq
where we use the same symbol to identify a chiral superfield or its scalar component.
The soft masses $m_{N,L}^2$, the trilinear ($A_N$) and bilinear ($B_N$) parameters are also $3\times 3$ matrices in generation space and all flavor indices are contracted with the usual rules of matrix product. All soft terms can be incorporated in the Lagrangian in \eq{susyL} by analytically continuing the parameters into superspace, with the replacement
\begin{eqnsystem}{replac}
\lambda_N &\to& (1+\theta^2 A_N ) \lambda_N ,\\
M&\to& (1+\theta^2 B_N)M ,\\
Z_a &\to& 1-\theta^2\bar\theta^2 {m}_a^2 ~~~~a=N,L,H_u  .
\end{eqnsystem}

The non-renormalization theorem of supersymmetry guarantees that the superpotential parameters $\lambda_N$ and $M$ are not modified by quantum corrections, while all the renormalization is contained only in the wave functions $Z_a$. Therefore the expression in eq.~(\ref{replac}c) is not stable under renormalization flow. Quantum effects generate corrections to $Z_a$ for all of its $\theta$ components, affecting the soft terms. In particular, the procedure of integrating out the right-handed neutrinos at the scale $M$, in the limit of exact supersymmetry, generates one-loop corrections to the wave functions of the $L$ and $H_u$ superfields which, in the $\overline {\rm MS}$ (or, equivalently, in the $\overline {\rm DR}$) subtraction 
scheme, are given by
\beq
\delta Z_L =\frac{\lambda_N^{R\dagger}}{16\pi^2}\left( 1-\ln \frac{M^{R\dagger}M^R}{\Lambda^2}\right) \lambda_N^R, ~~~~~~\delta Z_{H_u} ={\rm Tr}~  \delta Z_L,
\label{wr}
\eeq
\beq
\lambda_N^R \equiv {Z_N^{-1/2}}^T \lambda_N Z_L^{-1/2} Z_{H_u}^{-1/2}, ~~~~M^R\equiv {Z_N^{-1/2}}^T M Z_N^{-1/2}. 
\eeq
Here $\Lambda$ is the ultraviolet cutoff and $\lambda_N^R$ and $M^R$ are the usual renormalized parameters defined in a superfield basis in which the kinetic terms are canonical. The trace in \eq{wr} is taken over flavor indices.

We can use the supersymmetric expression for the wave-function renormalization in \eq{wr} to obtain the quantum corrections to the soft terms by promoting the parameters into superspace. Since the quantum corrections $\delta Z$ generate $\theta^2$ terms, it is convenient to redefine the superfields in order to bring the one-loop corrected wave functions back into their tree-level form, given in eq.~(\ref{replac}c). This can be done with the superfield redefinitions
\beq
L \to \left( 1- \frac{\left. \delta Z_L \right|_0}{2}\right) \left( 1-\theta^2\left. \delta Z_L \right|_{\theta^2} \right) L ,~~~
H_u \to \left( 1- \frac{\left. \delta Z_{H_u} \right|_0}{2}\right) \left( 1-\theta^2\left. \delta Z_{H_u} \right|_{\theta^2} \right) {H_u},
\label{lred}
\eeq 
where we have used the expansions
\beq
\delta Z_{L,H_u} = \left. \delta Z_{L,H_u} \right|_0 +\theta^2 \left. \delta Z_{L,H_u} \right|_{\theta^2}
+{\bar\theta}^2 \left. \delta Z_{L,H_u} \right|_{{\bar\theta}^2}
+\theta^2\bar\theta^2 \left. \delta Z_{L,H_u} \right|_{\theta^2\bar\theta^2}.
\eeq
Note that the transformation in \eq{lred} preserves the chiral properties of the superfields $L$ and $H_u$. This superfield rescaling transforms the kinetic terms in the following way
\beq
L^\dagger \left(1+\delta Z_L \right) L \to L^\dagger \left( 1+\theta^2\bar\theta^2 \left. \delta Z_L \right|_{\theta^2\bar\theta^2} \right) L,
\eeq
working in the one-loop approximation.
Comparing this expression (and the analogue for the $H_u$ superfield) with eq.~(\ref{replac}c), we obtain the one-loop corrections to the soft masses
\beq
 \delta {m}_L^2 = - \left. \delta Z_L \right|_{\theta^2\bar\theta^2}, ~~~~~~
 \delta {m}_{H_u}^2 = - \left. \delta Z_{H_u} \right|_{\theta^2\bar\theta^2}.
 \label{sof1}
 \eeq
 
 Once we apply the superfield redefinitions in \eq{lred} to the superpotential, we obtain two effects. First, the term $1-\delta Z|_0/2$ gives the usual renormalization of the superpotential coupling costants. Second, the term $1-\theta^2 \delta Z|_{\theta^2}$ generates trilinear and bilinear soft terms. Starting from the superpotential
 \beq
 \mathscr{W} =E
\lambda_{E}  L H_{\rm d} +
U\lambda_{U}   Q H_{\rm u}  +
D \lambda_{D}   Q H_{\rm d}  +
\mu ~H_{\rm u} H_{\rm d}  ,
\eeq
 the superfield redefinition induces one-loop corrections to the soft terms
\beq
 \mathscr{L}_{\rm soft} =E
\lambda_{E} A_{E} L H_{\rm d} +
U\lambda_{U} A_{U}  Q H_{\rm u}  +
D \lambda_{D} A_{D}  Q H_{\rm d}  +
B_{\mu} \mu ~H_{\rm u} H_{\rm d} +{\rm h.c.} ,
\eeq
\beq
\delta A_E = - \left. \delta Z_L \right|_{\theta^2} ~~~~~~
\delta A_U = - \identity \left. \delta Z_{H_u} \right|_{\theta^2} ~~~~~~
\delta A_D = 0 ~~~~~~
\delta B_\mu = -\left. \delta Z_{H_u} \right|_{\theta^2}.
\label{sof2}
\eeq
Here $\identity$ is the identity matrix in flavor space.

Equations (\ref{sof1}) and (\ref{sof2}) give the quantum corrections to the soft terms induced by integrating out the right-handed neutrinos in terms of the supersymmetric corrections to the wave function $\delta Z$. Since we are working in the one-loop approximation, we can obtain the explicit result by simply promoting  the expression of $\delta Z$ given in \eq{wr}
into superspace according to the  tree-level superspace continuation of eq.~(\ref{replac}).
In the basis in which the right-handed neutrino mass matrix $M$ is diagonal, real and positive, we obtain
\beq
\left( \delta A_E \right)_{ij} =\frac{{\lambda_N^*}_{k i}{\lambda_N}_{\ell j}}{16\pi^2}
\left[ {B_N}_{k \ell} F\left( \frac{M_k}{M_\ell} \right) - 
\left( \ln \frac{\Lambda^2}{M_k^2} +1\right) {A_N}_{k \ell} \right]\label{cpv_typeI}
\eeq
\beq
\delta A_U  = \identity ~{\rm Tr}~\delta A_E, ~~~~
\delta A_D =0,~~~~
\delta B_\mu = {\rm Tr}~\delta A_E
\label{spit1}
\eeq
\begin{eqnarray}
\left(\delta m_L^2\right)_{ij} &=&\frac{1}{16\pi^2}
\left\{ -\left(
  {\lambda_N^*}_{\ell i}{\lambda_N}_{kj} \frac{{m_N^2}_{k\ell}}{2}
+{\lambda_N^*}_{ki}{\lambda_N}_{\ell j} \frac{{m_N^2}_{\ell k}}{2}
 +{\lambda_N^*}_{k\ell}{\lambda_N}_{kj} \frac{{m_L^2}_{i\ell}}{2} 
\right. \right.   
\nonumber \\
&& \left.
+{\lambda_N^*}_{ki}{\lambda_N}_{k\ell} \frac{{m_L^2}_{\ell j}}{2}
+{\lambda_N^*}_{ki}{\lambda_N}_{kj} m_{H_u}^2 
+{\lambda_N^*}_{\ell i}{\lambda_N}_{sj} {A_N^*}_{k \ell} {A_N}_{k s} 
\right) \left( \ln \frac{\Lambda^2}{M_k^2} +1\right) 
\nonumber \\
&&+{\lambda_N^*}_{ki}{\lambda_N}_{\ell j}\left[
{m_N^2}_{k\ell} +
\frac{{m_N^2}_{\ell k}}{2}
\left( \frac{M_k}{M_\ell} + \frac{M_\ell}{M_k}\right) \right] 
F\left( \frac{M_k}{M_\ell} \right)
 \\
&&+\left(
{\lambda_N^*}_{k i}{\lambda_N}_{sj} {B_N^*}_{\ell k} {A_N}_{\ell s}
+{\lambda_N^*}_{s i}{\lambda_N}_{\ell j} {A_N^*}_{k s} {B_N}_{k \ell} 
\right)
F\left( \frac{M_k}{M_\ell} \right)
\nonumber  \\
&& +{\lambda_N^*}_{k i}{\lambda_N}_{\ell j} 
\left. \left[
{B_N^*}_{sk} {B_N}_{s \ell}  F\left( \frac{M_k}{M_\ell} \right)
- \frac{\left(
{B_N^*}_{sk} {B_N}_{s \ell}+{B_N^*}_{\ell s} {B_N}_{ks}\right)}{2}
G\left( \frac{M_k}{M_s},\frac{M_\ell}{M_s}\right) 
\right] \right\}\nonumber 
\end{eqnarray}
\beq
\delta m_{H_u}^2 = {\rm Tr}~ \delta m_L^2
\label{spit2}
\eeq
where the functions $F$ and $G$ are normalized such that $F(1)=G(1,1)=1$\footnote{Such loop functions arise when expanding in superspace
the logarithm of a matrix, according to the identity
\begin{eqnarray*}
\left[ \ln \Delta \left( 1+\theta^2 X +{\bar \theta}^2Y+\theta^2\bar\theta^2 W\right) \right]_{ij} &=&
\delta_{ij} \ln \Delta_i +\left(  \theta^2 X_{ij} + {\bar \theta}^2Y_{ij} +\theta^2\bar\theta^2 W_{ij}\right) f\left( \frac{\Delta_i}{\Delta_j} \right) +\\&&+\theta^2\bar\theta^2\left( X_{ik}Y_{kj}+Y_{ik}X_{kj}\right) g \left( \frac{\Delta_i}{\Delta_k}, \frac{\Delta_j}{\Delta_k} \right)
\end{eqnarray*}
valid for a diagonal matrix $\Delta$ and generic matrices $X$, $Y$, $W$, where 
$$f(x)=\frac{x\ln x}{x-1} ~~~~~~g(x,y)=\frac{x}{x-y}\left( \frac{\ln x}{x-1} - \frac{\ln y}{y-1}\right).$$}:
\beq
F(x)=\frac{2x\ln x}{x^2-1}, ~~~~G(x,y) = \frac{4xy}{x^2-y^2} \left( \frac{\ln x}{1-x^2}-\frac{\ln y}{1-y^2}\right).
\eeq
It is interesting to note how the loop functions emerge in the method of analytic continuation not as the result of integrals (like in ordinary Feynman diagram calculations), but rather as the expansion of a logarithmic function in superspace.

To make the result more transparent, we consider the degenerate case in which $m_{N,L,H_u}^2$ and $M$ are proportional to the identity (but $\lambda_N$, $A_N$ and $B_N$ are general matrices in flavor space). In this case, the expressions for $\delta A_E$ and for $\delta m_L^2$ become (in matrix notation)
\begin{eqnsystem}{sys:degenerate}
\delta A_E &=&\frac{\lambda_N^\dagger}{16\pi^2} \left[ B_N -A_N \left( \ln \frac{\Lambda^2}{M^2} +1\right) \right] \lambda_N ,\label{soff1}\\
\delta m_L^2 &=& \frac{\lambda_N^\dagger}{16\pi^2} \bigg[ 2m_N^2- \bigg( m_N^2+m_L^2+m_{H_u}^2+A_N^\dagger A_N \bigg) \bigg( \ln \frac{\Lambda^2}{M^2} +1\bigg)   \nonumber \\ 
&&+B_N^\dagger A_N +A_N^\dagger B_N +\frac{\left( B_N^\dagger B_N - B_N B_N^\dagger \right)}{2} \bigg] \lambda_N ,
\label{soff2}
\end{eqnsystem}
while the other soft terms are still given by eqs.~(\ref{spit1}) and (\ref{spit2}).
In the case in which the matrix $B_N$ is universal or hermitian, the last term vanishes.

The logarithmic terms in \ eqs.~(\ref{sys:degenerate}) reproduce the renormalization-group evolution of 
the soft terms from $\Lambda$ to the mass scale $M$ of the right-handed neutrinos, once the leading logs 
are resummed with the usual technique. Our expressions also give the finite threshold corrections, 
confirming and extending the results in ref.~\cite{FP}.

The terms proportional to $B_N$ in eqs.~(\ref{soff1}) and (\ref{soff2}) are  (scheme-independent) finite 
threshold corrections at the scale $M$. They are completely analogous to the gauge-mediation effects 
in presence of superpotential couplings between messengers and matter, since $N$ is playing 
the role of a messenger field coupled to a spurion superfield $X=M+\theta^2 F$ with $F=B_NM$.
Just as in that case, as long as $[X,X^\dagger]=0$, there are no one-loop contributions to the square 
scalar masses proportional to $F^2/M^2=B_N^2$~\cite{Giudice:1998bp}. This is because the one-loop 
correction to the wave function is proportional to $\ln X^\dagger X$, which can be written as 
$\ln X +\ln X^\dagger$, if $X$ and $X^\dagger$ commute; no $\theta^2\bar\theta^2$ term can then be present. 
For this reason, in the basis where $M$ is diagonal, the contribution in \eq{soff2} is proportional to 
$[B,B^\dagger]$. Our result for the threshold correction to the soft masses proportional to 
Re$(A_NB_N^\dagger)$ disagrees with ref.~\cite{Farzan:2003gn} in the sign, and the term proportional 
to~$m^2$ agrees with ref.~\cite{FP}.

\smallskip

The novelty of our results with respect to previous analyses lies primarily in the new contributions to 
$A_U$ and $B_{\mu}$. Note that, differently than in the case of $\delta A_E$, the corrections to $A_U$ 
and $B_{\mu}$ are proportional to a trace involving $\lambda_N^\dagger \lambda_N$. 
Therefore, while $A_E$ is sensitive to the flavor structure, the corrections to $A_U$ and $B_{\mu}$ 
are sizable whenever any of the entries of $\lambda_N$ is large, irrespectively of its flavor index.

\subsection{Type-III supersymmetric see-saw}

An alternative see-saw sector is described by weak SU(2) triplet chiral superfields $N^a$ ($a=1,2,3$ is 
the SU(2) index) with zero hypercharge and with superpotential
\beq\label{eq:Lseesaw3}
\mathscr{W} = N^{aT}\frac{M}{2} {N^a} +  N^{aT}\lambda_N L \tau^a H_{\rm u}
\eeq
such that the formula for neutrino mass matrix remains the same as in type-I. As in the previous case, 
there are 3 generations of $N^a$ and the parameters $M$ and $\lambda_N$ are $3\times 3$ matrices in flavor 
space, with flavor indices omitted. The soft terms are given by
\beq
\mathscr{L}_{\rm soft} =
-N^{a\dagger} m_N^2  N^a -L^\dagger m_L^2  L- {H_{\rm u}}^\dagger m_{H_{\rm u}}^2{H_{\rm u}}+\left( N^{aT} \frac{B_NM}{2}N^a+N^{aT} A_N \lambda_N  L\tau^a H_{\rm u}  +{\rm h.c.}\right) 
\eeq
The calculation of the corrections to the low-energy soft terms is completely analogous to
the previous case, and the supersymmetric wave function renormalization is still given by \eq{wr},
after multiplying the right-hand side by a factor of 3. This extra factor of 3 counts the states 
circulating in the loop of the wave-function diagram. Hence, all soft terms are given by the 
expressions derived for type-I see-saw, multiplied by a factor of 3.

There is one important  novelty of type-III see-saw, related to the fact that the 
heavy fields $N^a$ now have gauge interactions, such that also the vectors supermultiplet 
gets a wave-function renormalization that depends on $M$.  This is the well known correction
to the gauge $\beta$ function, which, after analytical continuation into superspace, also
implies a  finite one-loop corrections at the threshold $M$ to the gaugino mass. This
contribution is completely analogous to the familiar case of gauge mediation, and it is
proportional to the discontinuity of the corresponding $\beta$-function at the scale $M$.
The wino mass $M_2$ receives a correction
\beq
\delta M_2 = \frac{g^2}{8\pi^2}{\rm Tr}~B_{N} ,
\label{cpv_typeIII}
\eeq
while the gluino and bino masses, $M_3$ and $M_1$,  are unaffected because the superfield $N^a$ 
carries neither color nor hypercharge. Since the effect is proportional to the complex parameter
$B_N$, this correction introduces a new CP-violating phase in the gaugino sector.

\subsection{Type-II supersymmetric see-saw}

Another possible see-saw sector is obtained by introducing weak triplets $T$ and $\bar T$
with hypercharge $Y_T =-Y_{\bar T}= -2Y_L$, such that the relevant superpotential is
\beq \mathscr{W} =  M\, T\bar{T} +\frac{1}{2}\left(\lambda_T\, L^T \varepsilon \tau^a L T^a+\lambda_{H_{\rm u}}
H_{\rm u}^T\varepsilon \tau^a H_{\rm u} \bar T^a +
 \lambda_{H_{\rm d}} 
H_{\rm d}^T\varepsilon \tau^a H_{\rm d}  T^a \right) .
\eeq
Here $\lambda_T$ is a $3\times 3$ matrix in flavor space, and $\varepsilon=i\tau^2$ is the anti-symmetric tensor.
One single $T$ is enough to induce generic neutrino masses. Since $\bar T$ does not couple to leptons, the 
neutrino mass matrix is $m_\nu^{ij} = \lambda_T^{ij} \lambda_{H_{\rm u}} {\langle H_{\rm u}\rangle}^2/M$ and 
its flavor structure is identical to the one of the Yukawa coupling $\lambda_T$. 

The soft terms can be introduced by analytically continuing the parameter into superspace according to the rules
\begin{eqnsystem}{replac3}
Z_a \to 1-\theta^2\bar\theta^2 {m}_a^2 ~~~~a=T,{\bar T},L,H_u,H_d\hspace{3cm}  \\
\lambda_T \to (1+\theta^2 A_{\lambda_T} ) \lambda_T ,
~~~~ \lambda_{H_{u,d}} \to (1+\theta^2 A_{\lambda_{H_{u,d}}} ) \lambda_{H_{u,d}} ,
 ~~~~ M\to (1+\theta^2 B_T)M .
\end{eqnsystem}
For simplicity, we assume flavor universality for the soft terms and thus $m_L^2$ and $A_{\lambda_T}$ in eq.~(\ref{replac3}) do not carry flavor indices.

In this case we find the following one-loop corrections for the supersymmetric wave functions
of the fields $L$, $H_{\rm u}$, and $H_{\rm d}$
\begin{eqnsystem}{sys:3}
\delta Z_L  &=&3\frac{\lambda_T^\dagger \lambda_T}{16\pi^2}
\left( \ln \frac{\Lambda^2}{M^\dagger M} +1\right) ,\\
\delta Z_{H_{\rm u}} &=&3\frac{\lambda_{H_{\rm u}}^\dagger \lambda_{H_{\rm u}}}{16\pi^2} \left( \ln \frac{\Lambda^2}{M^\dagger M} +1\right) ,\\
\delta Z_{H_{\rm d}} &=&3\frac{\lambda_{H_{\rm d}}^\dagger \lambda_{H_{\rm d}}}{16\pi^2} \left( \ln \frac{\Lambda^2}{M^\dagger M} +1\right) .
\end{eqnsystem}
Promoting the wave functions into superspace with the prescription given in eq.~(\ref{replac3}), 
we find
\beq
\delta A_E  =3\frac{\lambda_T^\dagger \lambda_T}{16\pi^2}
\left[ B_T - A_{\lambda_T} 
\left( \ln \frac{\Lambda^2}{M^2} +1\right)  \right] +\identity ~F_D,
\eeq
\beq
\delta A_U = \identity ~F_U,~~~~
\delta A_D = \identity ~F_D,~~~~
\delta B_\mu = F_U +F_D,
\eeq
\beq
F_{U,D} = 3\frac{\left| \lambda_{H_{u,d}}\right|^2}{16\pi^2}
\left[ B_T - A_{\lambda_{H_{u,d}}} 
\left( \ln \frac{\Lambda^2}{M^2} +1\right)  \right] ,
\eeq
\beq
\delta m_L^2 = 3\frac{\lambda_T^\dagger \lambda_T}{16\pi^2} \left[ m_T^2 +m_{\bar T}^2 -\left( 2 m_L^2 +m_T^2 + \left| A_{\lambda_T} \right|^2 \right) \left( \ln \frac{\Lambda^2}{M^2} +1\right) +2~{\rm Re}\left( A_{\lambda_T}^*B_T \right) \right] ,
\eeq
\beq
\delta m_{H_u}^2 = 3\frac{\left| \lambda_{H_u}\right|^2}{16\pi^2} \left[ m_T^2 +m_{\bar T}^2 -\left( 2 m_{H_u}^2 +m_{\bar T}^2 + \left| A_{\lambda_{H_u}} \right|^2 \right) \left( \ln \frac{\Lambda^2}{M^2} +1\right) +2~{\rm Re}\left( A_{\lambda_{H_u}}^*B_T \right) \right] ,
\eeq
\beq
\delta m_{H_d}^2 = 3\frac{\left| \lambda_{H_d}\right|^2}{16\pi^2} \left[ m_T^2 +m_{\bar T}^2 -\left( 2 m_{H_d}^2 +m_{T}^2 + \left| A_{\lambda_{H_d}} \right|^2 \right) \left( \ln \frac{\Lambda^2}{M^2} +1\right) +2~{\rm Re}\left( A_{\lambda_{H_d}}^*B_T \right) \right] .
\eeq

In type-II see-saw the threshold corrections at the scale $M$ contribute, in the squark
sector, to both $A_U$ and $A_D$. Moreover, since the fields $T$ and $\bar T$ carry both
weak and hypercharge quantum numbers, we find corrections to both wino and bino masses~\cite{Goto:2010ce}:
\beq\label{eq:M1M2}
\delta M_2 = \frac{g^2}{4\pi^2}B_T ,~~~~
\delta M_1 = \frac{3g^{\prime 2}}{8\pi^2}B_T .
\label{cpv_typeII}
\eeq
All formul\ae{} are trivially extended to the cases of more $T+\bar T$ repetitions or hybrids
between type-I and/or type-II and/or type-III see-saws. RGE-improving and RGE running from
$M$ down to the weak scale is well known.

\section{Electric dipole moments}\label{EDM}

In the MSSM, the potential sources of CP violation are the $\mu$ term, the gaugino masses 
$M_{i}$ (with $i=1,2,3$), $B_{\mu}$ and the $A$ terms $A_F$ (with $F=E,U,D$) as well as the 
sfermion masses.
As well known, only some 
of their combinations provide physical CP violating phases~\cite{dgh}. 
As shown in the previous section, complex $A_N$- and $B_N$-terms of the see-saw heavy
fields give corrections to the $B_{\mu}$ term and to some $A_F$-terms of the light MSSM
fields\footnote{We here neglect CP violation in the neutrino
Yukawa couplings. We remind that, if the neutrino Yukawa couplings $\lambda_N$ are the
only source of CP violation, the induced leptonic EDM are small~\cite{edminseesaw,FP},
below the future planned experimental sensitivities~\cite{exp_edm}. The situation is
different in SUSY GUT see-saw models~\cite{gutedm,Hisano:2008hn}.}.
To reach the usual phase convention $B_\mu\mu>0$ 
(i.e.\ $\tan\beta > 0$) a complex $B_\mu$ needs an
opposite phase in $\mu$.


%

Specializing the well known generic formul\ae{} for EDM from sparticle loops to this case,
and assuming sparticles degenerate at a common mass $m_{\rm SUSY}$, the electron EDM is
\begin{eqnsystem}{sys:EDM}
{d_e}
&=&
-\frac{m_ee}{4\pi m_{\rm SUSY}^2}
\left[
\left(\frac{5\alpha_2}{24}
\sin\phi_{2}
+\frac{\alpha_Y}{24}
\sin\phi_{1}
\right)
\tan\beta
+\frac{\alpha_Y}{12}
\sin\phi_{A_e}
\right]~,
\label{mssmde}\\
\riga{the quark EDM are (neglecting the ${\cal O}(\alpha_Y)$ contribution):}\\
%
{d_d}
&=&
-\frac{m_de}{{4\pi}m_{\rm SUSY}^2}\left[
\left(
\frac{2\alpha_3}{27}\sin\phi_{3}
+\frac{7\alpha_2}{24}\sin\phi_{2}
\right)\tan\beta
+\frac{2\alpha_3}{27}
\sin\phi_{A_d}
\right]
,
\\
{d_u}
&=&
+\frac{m_ue}{4\pi m_{\rm SUSY}^2}
\left[
\left(
\frac{4\alpha_3}{27}\sin\phi_{3}
+\frac{\alpha_2}{4}\sin\phi_{2}
\right)\tan^{-1}\beta
+
\frac{4\alpha_3}{27}\sin\phi_{A_u}
\right]
,
\\[2mm]
\riga{and their chromo-electric dipoles are:}\\[1mm]
d_d^c
&=&
-\frac{m_d}{{4\pi}m_{\rm SUSY}^2}
\left[
\left(
\frac{5\alpha_3}{18}\sin\phi_{3}
+\frac{\alpha_2}{8}\sin\phi_{2}
\right)
\tan\beta
+
\frac{5\alpha_3}{18}\sin\phi_{A_d} \right],\\
d_u^c
&=&
-\frac{m_u}{{4\pi}m_{\rm SUSY}^2}
\left[
\left(
\frac{5\alpha_3}{18}\sin\phi_{3}
+\frac{\alpha_2}{8}\sin\phi_{2}
\right)
\tan^{-1}\beta
+
\frac{5\alpha_3}{18}\sin\phi_{A_u}
\right],
\label{mssmdq}
\end{eqnsystem}
where the quark masses and all parameters are renormalized at the scale $m_{\rm SUSY}$ and
\beq \phi_{i}={\rm arg}\,(\mu M_{i}),\qquad
\phi_{A_e}={\rm arg}\,(M_{1}A^{*}_{e}),\qquad
\phi_{A_q}={\rm arg}\,(M_{3}A^{*}_{q})\eeq 
in the convention of~\cite{Pokorski:1999hz}.
%
%
%
%
%
The neutron EDM $d_n$ can be estimated from the naive quark model as $d_n\approx\frac{4}{3} d_d - \frac{1}{3} d_u$.
A better estimate is obtained from QCD sum rules~\cite{Pospelov:1999ha,Pospelov:2000bw} taking
also into account QCD renormalization of $d_q$ and $d_q^c$ from $m_{\rm SUSY}\approx m_t$ down
to $m_n$~\cite{Degrassi:2005zd}:
\begin{equation}
d_n =  1.2 d_d - 0.3 d_u + 0.2 e\, d^c_d + 0.8 e \,d^c_u \,
\label{Eq:dn_odd}
\end{equation}
up to an uncertainty of about $\pm50\%$. Given the current experimental bounds
$|d_n| < 2.9 \times 10^{-26}~e\,\rm{cm}~(90\% \rm{CL})$~\cite{Baker:2006ts} and
$|d_{\rm Hg}| < 3.1 \times 10^{-29}~e\,\rm{cm}~(95\% \rm{CL})$~\cite{Griffith:2009zz},
it might be that $|d_{\rm Hg}|$ is more sensitive than $d_n$ to SUSY effects~\cite{Pospelov:1999ha,Pospelov:2000bw}.
Given the large theoretical uncertainties affecting the prediction of $d_{\rm Hg}$,
we prefer here to be conservative and focus only on $d_n$.

\bigskip

Before proceeding, we recall the correlation between $d_e$ and $a_{\mu}=(g-2)_{\mu}/2$.
Recent analyses of $a_{\mu}$ point towards a $3\sigma$ discrepancy~\cite{g_2_th,passera_mh}:
$\Delta a_{\mu}\!=\!a_{\mu}^{\rm exp}\!-\!a_{\mu}^{\rm SM}\approx(3\pm 1)\times 10^{-9}$. Therefore, it is interesting to evaluate the expected value for $d_e$ within SUSY
see-saw models assuming that the above discrepancy is due to supersymmetry.
Assuming again sparticles degenerate at a common mass $m_{\rm SUSY}$, we find
\begin{equation}
{|d_e|} \approx
10^{-27}{e\,\rm{cm}}\times\frac{\Delta a_{\mu}}{3\times 10^{-9}}
\frac{|\phi_{2}|}{10^{-3}}\,.
\label{edm_gm2}
\end{equation}
We recall that  $|\phi_2|\equiv |{\rm arg}\,(\mu M_{2})| \approx 10^{-3}$
is the natural size of a loop induced CP violating phase, as arising in the context
of type-II and III SUSY see-saw models.



\subsection{Type-I SUSY-see-saw}
Complex $A_N$ and $B_N$ terms induce two main effects: \begin{itemize}
\item[i)] contributions to the $A$-terms of the electron and of the up-quark:
$\delta A_e\sim (\lambda^{\dagger}_{N}\lambda_{N})_{ee} (A_N, B_N)$ and
$\delta  A_u \sim{\rm Tr}(\lambda^{\dagger}_{N}\lambda_{N})(A_N, B_N)$;
\item[ii)] a contribution to
$ B_{\mu}$: $\delta B_{\mu}\sim{\rm Tr}(\lambda^{\dagger}_{N}\lambda_{N})(A_N,B_N)$.
Going to the standard phase convention where $B_{\mu}\mu$ is real and positive, this 
needs a complex $\mu$ term.\footnote{Even within the CMSSM a complex $A_0$ indirectly
induces a complex $\mu$, and thus studies of EDM from complex $A$-terms and real $\mu$
term do not seem motivated.}
A qualitatively similar result holds in the next-to-minimal MSSM, where the $\mu$ term
is replaced by the vev of a singlet.

\end{itemize}
Therefore, type-I see-saw with complex $A_N$ or $B_N$ predict contributions to the EDM
typical of a complex $\mu$-term. In fact, the effects driven by a complex $\mu$ to the
electron EDM are parametrically enhanced by a factor of $(\alpha_2/\alpha_Y)\tan\beta$
compared to the effects from a complex $A$-term, see eq.~(\ref{mssmde}).
In the quark sector, the enhancement is only of order $\tan\beta$ as the gluino effects
are dominant both in case of a complex $A_u$ or $\mu$ terms, see eq.~(\ref{mssmdq}).

A characteristic prediction of this scenario is the ratio $d_e/d_n$, that can be
computed knowing the sparticle spectrum. The $\mu$ term with a small complex phase
gives $\phi_{3}=\phi_{2}=\phi_{1}\equiv \phi_i$, so that
\beq 
|d_n| \approx 20 |d_e|  \approx
2 \times 10^{-27}e\,{\rm cm} \frac{\tan\beta }{10}
\left(\frac{200\,\rm{GeV}}{m_{\rm SUSY}}\right)^2
\frac{|\,\phi_i\,|}{10^{-4}}\ .
\eeq
The ratio $|d_n/d_e|\approx 20$ gets reduced if squarks are heavier than sleptons,
and can be computed once sparticles are discovered and their masses measured.

\medskip

As an illustrative example of a full computation, we assume now a constrained-MSSM (CMSSM)
spectrum. Following~\cite{GRS}, we perform a `naturalness' scan of the parameter space, i.e.\ 
with density inversely proportional to the fine-tuning, such that regions with
higher points density are proportionally more likely.
Operatively, this is obtained by sampling the CMSSM parameters at the GUT scale as
\beq
m_0 = [0..1] \tilde{m}, \qquad M_{1/2} = [0...1] \tilde{m},\qquad
A_0 = [-1...1] \tilde{m},\qquad \mu = [-1...1]\tilde{m}
\label{natural_scan}
\eeq
where $[a...b]$ denotes a random number between $a$ and $b$, and fixing the overall
scale $\tilde{m}$ by requiring the correct scale of electro-weak symmetry breaking,
$M_Z^2\simeq -2m_{H_{\rm u}}^2-2 |\mu|^2$ at tree level for large $\tan\beta$. Most
of the obtained sparticle spectra have sparticle masses below $M_Z$ and are thereby
excluded by the current constraints which make the CMSSM fine-tuned, so we discard 
them.

\begin{figure}[t]
\begin{center}
\includegraphics[width=0.47\textwidth]{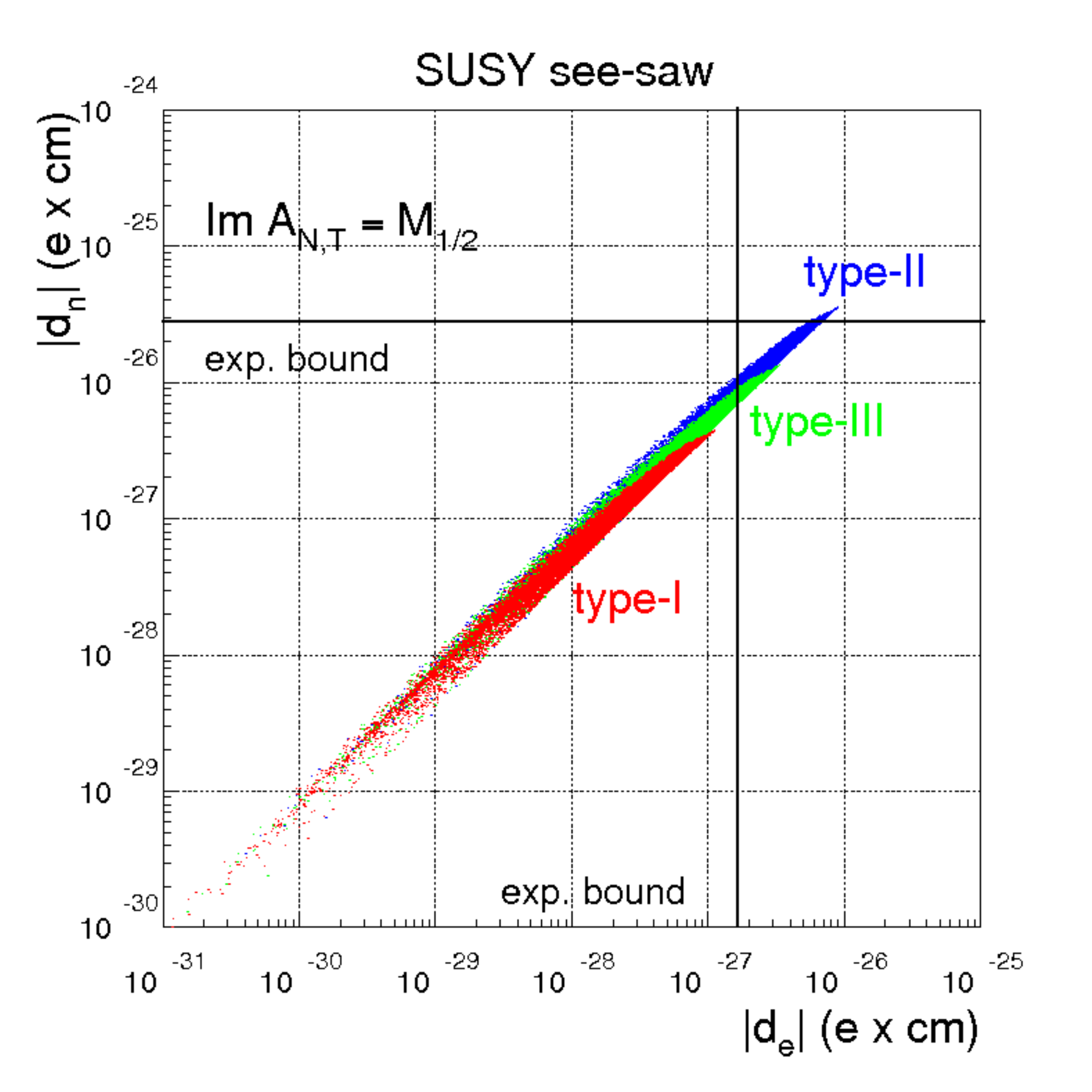}\qquad
\includegraphics[width=0.47\textwidth]{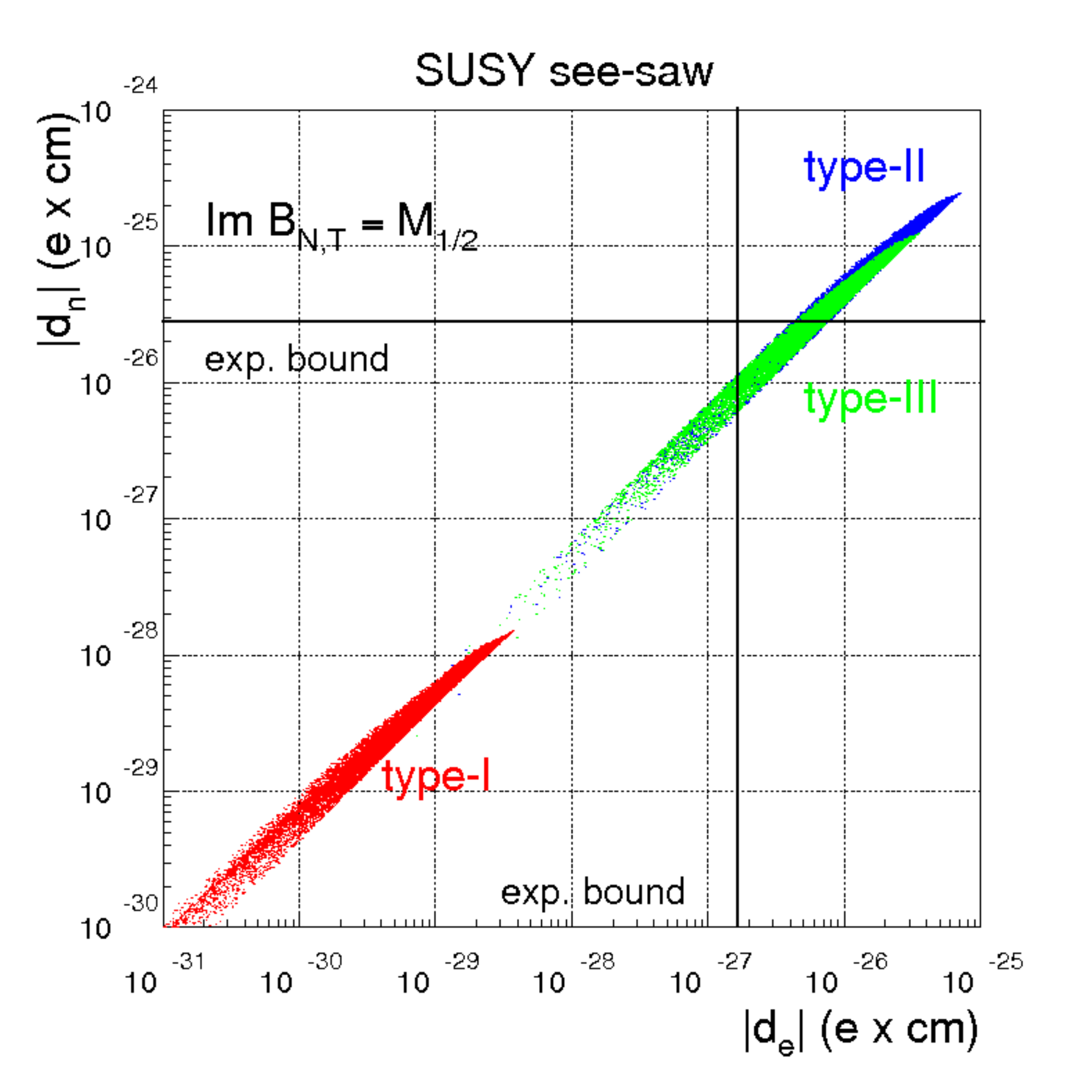}
\caption{{\em Predictions of SUSY-see-saw models for $|d_n|$ versus $|d_e|$.
We assume a CMSSM-like spectrum making a `naturalness' scan of the parameter
space as described in eq.~(\ref{natural_scan}), such that higher density of 
points means higher probability.
We assume the degenerate case, in which $m_{N,L,H_u}^2$, $A_{N,T}$, $B_{N,T}$ and 
$M$ are proportional to the identity, and we set} $\tan\beta = 10$,
${\rm Tr}(\lambda^{\dagger}_{N}\lambda_{N})={\rm Tr}(\lambda^{\dagger}_{T}\lambda_{T})=
|\lambda_{H_{\rm u}}|^2 = |\lambda_{H_{\rm d}}|^2 = 10^{-3}$
and $M=10^{12}\GeV$ such that observed neutrino masses are typically reproduced
and leptogenesis is possible in all see-saw models. For definiteness we assumed
${\rm Im}\, A_{N,T}=M_{1/2}$ (left) and ${\rm Im}\, B_{N,T}=M_{1/2}$ (right).}
\label{fig:scan}
\end{center}
\end{figure}

\medskip

As an illustrative example, in our numerical analysis we consider the degenerate case
in which $m_{N,L,H_u}^2$, $A_{N,T}$, $B_{N,T}$ and $M$ are proportional to the identity.
Keeping the parameter $\tan\beta$ fixed to $\tan\beta=10$, and assuming
${\rm Tr}(\lambda^{\dagger}_{N}\lambda_{N})={\rm Tr}(\lambda^{\dagger}_{T}\lambda_{T})=
|\lambda_{H_u}|^2=|\lambda_{H_d}|^2=10^{-3}$,
in fig.~\ref{fig:scan} on the left (right) we show the predictions for $|d_n|$ vs.\ $|d_e|$
in case of ${\rm Im}\,A_{N,T} = M_{1/2}$ (${\rm Im}\,B_{N,T}=M_{1/2}$), where $M_{1/2}$
is the unified gaugino mass parameter.
Both EDM roughly scale as ${\rm Im}\,(A,B)\times\tan\beta\times {\rm Tr}(\lambda^{\dagger}_{N,T}\lambda_{N,T})$.
As shown in fig.~\ref{fig:scan}, the case where ${\rm Im}\,A_{N,T} \neq 0$ predicts EDM
enhanced by a large log factor $\log\Lambda^2/M^2$ compared to the case where
${\rm Im}\,B_{N,T} \neq 0$ (see also eq.~(\ref{cpv_typeI}))\footnote{If
we assume a fixed ${\rm Im}\,(A_{N,T},B_{N,T})$ independent of $\tan\beta$, $d_e$ and $d_n$ 
have an extra $\tan\beta$ enhancement, because a large $\tan\beta$ is obtained from a
small $B_{\mu}\simeq m^2_A/\mu\tan\beta$, consequently CP-violating phases would scale
as $\phi_2\propto \tan\beta$.}. As we can see, $|d_n|$ and $|d_e|$ are highly correlated
such that $|d_n|\approx 10 |d_e|$ in most of the CMSSM parameter space. The ratio $d_n/d_e$ 
depends on the specific sparticle mass spectrum considered.

\bigskip

Our results show that contributions to the EDMs require phases in $A_{N,T}$ or $B_{N,T}$ but
not necessarily in both. Therefore even type-I models with vanishingly small $B_N$,
like those motivated by soft leptogenesis~\cite{Grossman:2003jv,D'Ambrosio:2003wy}, can lead
to observable effects.

\subsection{Type-III SUSY-see-saw}

The effects driven by  $A_N $ are similar to type-I see-saw up to order one factors, as
shown by fig.~\ref{fig:scan} on the left. The qualitatively new feature is the correction
to $M_2 \sim g_2^2 B_M$. This effect is dominant as long as $g_2 \gg \lambda$ i.e.\ $M_{N_{1,2,3}}\circa{<}10^{14}\GeV$, and we consider this new scenario.

We recall that a common phase of the gaugino masses $M_{1,2,3}$ is usually turned into a
common phase of the $A$-terms via an $R$-transformation. This is not possible if $M_{1,2,3}$
have different complex phases. Thereby it is more convenient to keep $M_2$ complex.

Under the usual RGE running a complex $M_2$ induces complex $A$-terms and a complex
$B_{\mu}$ (thus a complex $\mu$ in the basis where $B_\mu\mu$ is real) that we systematically
take into account in our numerical analysis.

\medskip

Fig.~\ref{fig:scan} shows the typical range of $d_e$ and $d_n$ again assuming
${\rm Tr}(\lambda^{\dagger}_{N}\lambda_{N})={\rm Tr}(\lambda^{\dagger}_{T}\lambda_{T})=
|\lambda_{H_{\rm u}}|^2 = |\lambda_{H_{\rm d}}|^2 = 10^{-3}$
and ${\rm Im}(A_{N,T}, B_{N,T}) = M_{1/2}$. As in type-I SUSY see-saw, it turns out
that $|d_n|\approx 10 |d_e|$. The effect of $A_{N,T}$ is quantitatively similar to
the type-I see-saw case, while the effect of $B_{N,T}$ can now be much larger, in
view of the gauge threshold effect.

\subsection{Type-II SUSY-see-saw}
Type II see-saw allows to predict the flavor of the Yukawa matrix $\lambda_T$ in terms of
the neutrino mass matrix $m_\nu$, but the overall size of $\lambda_T$ is not fixed in terms
of $m_\nu$ and $M$, as the ratio between $\lambda_T$ and $\lambda_{H_{\rm u}}$ is unknown.
It is possible to make arbitrarily small either $\delta A_E$ (that contributes to the
electron EDM) or $\delta A_U$ (that contributes to the neutron EDM) but not both.

As in type-III see-saw there is an extra effect due to the gauge couplings, see \eq{eq:M1M2}.
All electroweak gauginos are now affected, but the new effect due to the bino is not dominant. Thereby the gauge-induced EDM are qualitatively similar
to type-III and it turns out that $|d_{e,n}|_{\rm II} \approx 2 |d_{e,n}|_{\rm III}$
as shown analytically by eqs.~(\ref{cpv_typeIII}), (\ref{cpv_typeII}) and numerically
in fig.~\ref{fig:scan} on the right.

If CP violation is driven by $A_T $, the predictions for the EDM highly depend on
the values of the relevant Yukawa couplings. Assuming 
${\rm Tr}(\lambda^{\dagger}_{N}\lambda_{N})={\rm Tr}(\lambda^{\dagger}_{T}\lambda_{T})= |\lambda_{H_{\rm u}}|^2 = |\lambda_{H_{\rm d}}|^2 = 10^{-3}$, all the types of see-saw
models provide the same predictions for the EDM up to order one factors, as shown in
fig.~\ref{fig:scan} on the left.

\section{Conclusions}
\label{end}

We considered the corrections to soft terms that arise at one loop in supersymmetric models
when integrating out heavy field related to the see-saw sector. The use of analytic continuation
in superspace makes clear that the possibly complex $B$-term of the heavy field propagates
to the $A$-terms and $B$-terms of all light fields coupled to it. Furthermore, the possibly 
complex $A$-terms of heavy fields have a similar effect, already from the well-known RGE.

In particular, the heavy fields present in see-saw models to mediate the neutrino mass 
operator $(LH_{\rm u})^2$ necessarily couple to $L$ and to $H_{\rm u}$, making complex 
not only the $A$-terms that involve the leptons $L$, but also those involving $H_{\rm u}$ 
(coupled to up-quarks), as well as its $B_{\mu}$ term.
In the standard phase convention where $B_\mu \mu$ is real, the $\mu$-term has a phase
opposite to $B_\mu$. Consequently see-saw models give contributions to the electron EDM
and to the quark EDM that were previously neglected. Such effects are proportional to
the unknown neutrino Yukawa couplings squared.

\smallskip

Furthermore, within type-II and type-III see-saw models, where neutrino masses are 
mediated by heavy triplets under $\SU(2)_L$, there are new effects proportional to 
their well-known gauge couplings squared: at one loop the $B_N$-term makes complex the
weak gaugino masses $M_2$ and (in type-II only) $M_1$. Again this induces electric 
dipole moments for all fermions, not only leptons. A order unity complex phase in the 
see-saw $B$ term gives a small phase, of order $\alpha_2/4\pi$, to light soft terms,
and consequently EDM just below the present bounds, if sparticles exist around the 
weak scale.

Lepton EDM are predicted to be proportional to the mass of the corresponding lepton,
but nevertheless $d_e$ is presently a better probe than $d_\mu$ and especially $d_\tau$.
Moreover, there is a loose connection between $d_e$ and lepton flavor violating processes 
like $\mu\to e\gamma$: in type-I see-saw $d_e$ and $\mu\to e\gamma$ are both generated
by the neutrino Yukawa couplings, whose flavor structure is however unknown, while in
type-II and III the presumably dominant gauge threshold effect only gives rise to electric 
dipoles.

\medskip

In conclusion, there are previously unnoticed signals of SUSY-see-saw models which are 
not confined to leptons. On the contrary, SUSY see-saw models can induce observable 
electric-dipole moments for the electron and for the neutron, in a characteristic ratio.

\paragraph{Acknowledgments:} We thank Michele Frigerio, Riccardo Rattazzi, Andrea Romanino 
for useful discussions.
P.P. thanks the CERN where part of his work was carried out.
The work of P.P. was supported in part by the Cluster of Excellence ``Origin
and Structure of the Universe'' and by the German Bundesministerium f{\"u}r Bildung
und Forschung under contract 05H09WOE.

\end{document}